\pdfoutput=1
\documentclass[pre,aps]{revtex4}

\usepackage{times}
\usepackage{amsmath}
\usepackage{graphicx}
\usepackage{amssymb}
\usepackage{color}

\begin{document}

\title{Solitary Waves in a Discrete Nonlinear Dirac equation}

\author{Jes\'us Cuevas--Maraver}
\affiliation{Grupo de F\'{\i}sica No Lineal, Departamento de F\'{\i}sica Aplicada I, Universidad de Sevilla. Escuela Polit{\'e}cnica Superior, C/ Virgen de \'Africa, 7, 41011-Sevilla, Spain\\
Instituto de Matem\'aticas de la Universidad de Sevilla (IMUS). Edificio Celestino Mutis. Avda. Reina Mercedes s/n, 41012-Sevilla, Spain}

\author{Panayotis\ G.\ Kevrekidis}
\affiliation{Department of Mathematics and Statistics, University of
Massachusetts, Amherst, MA 01003-9305, USA}

\author{Avadh Saxena}
\affiliation{Center for Nonlinear Studies and Theoretical Division, Los Alamos National Laboratory, Los Alamos, New Mexico 87545, USA}


\begin{abstract}
In the present work, we introduce a discrete formulation of the
nonlinear Dirac equation in the form of a discretization of
the Gross-Neveu model. The motivation for this discrete model proposal
is both computational (near the continuum limit) and theoretical
(using the understanding of the anti-continuum limit of vanishing
coupling). Numerous unexpected features are identified including
a staggered solitary pattern emerging from a single site excitation,
as well as two- and three-site excitations playing a role analogous
to one- and two-site, respectively, excitations of the discrete
nonlinear Schr{\"o}dinger analogue of the model. Stability exchanges
between the two- and three-site states
are identified, as well as instabilities that
appear to be persistent over the coupling strength $\epsilon$,
for a subcritical value of the propagation constant $\Lambda$.
Variations of the propagation
constant, coupling parameter and nonlinearity exponent are all
examined in terms of their existence and stability implications and
long dynamical simulations are used to unravel the evolutionary
phenomenology of the system (when unstable).
\end{abstract}
\maketitle

\section{Introduction}

Nonlinear dispersive waves  in physical systems
are often described by the nonlinear
Schr{\"o}dinger equation (NLSE), which is both mathematically and physically
studied in a broad range of settings including atomic
physics~\cite{emergent}, nonlinear optics~\cite{kivshar} and mathematical
physics~\cite{sulem,ablowitz}. Both the continuum and the
discrete~\cite{pgk,pelinov} installment
of the equation have been analyzed in detail. A principal
focus of the relevant properties, aside from issues of self-focusing
and wave collapse~\cite{sulem} has been the study of the
existence, stability and dynamics of solitary waves in this model,
both in lower-dimensional settings (such as one-dimensional solitons and multi-solitons)
and in higher dimensional settings (vortices, vortex rings, and related
structures).

However, recent years have seen a gradual increase of interest in the
study of near-relativistic settings, where a suitable generalization/extension
of the NLSE is the so-called nonlinear Dirac equation (NLDE)~\cite{gavri}.
In fact, a modified form of the NLSE (with additional terms)
is a special case limit of the NLDE at the low-energy limit as has been
demonstrated in~\cite{nogami}. Different realizations of the
NLDE have been proposed in the realm of high energy physics, including
the so-called massive Gross-Neveu model~\cite{gn3}, as well as the
massive Thirring model~\cite{t4}. Importantly, the equation has seen
a significant volume of studies from a more mathematical perspective.
Various aspects have been examined in this context, including
the spectral stability and the potential emergence of point spectrum
eigenvalues with nonzero real part (which has been shown to be
impossible to happen beyond the
so-called embedded thresholds)~\cite{boucom}, the orbital and asymptotic
stability under a series of relevant assumptions~\cite{boucuc}, the
nonlinear Schr{\"o}dinger (non-relativistic) limit and its instability
for nonlinearities beyond a critical exponent~\cite{comgus}, as well
as classical (Vakhitov-Kolokolov) and more suitable to this setting
(energy based) criteria~\cite{berko} for the linear stability of solitary
waves in the NLDE. A series of more computationally/physically
oriented studies both in the context of the stability/dynamics
of the NLDE solitary waves~\cite{cooper,niur_recent}
(again, in principle for arbitrary nonlinearity powers) and in that of
these structures in the presence of external fields~\cite{niurka}
have also recently appeared.

It would be relevant to mention, at least in passing, one more
framework where Dirac-type equations have received significant
attention in recent years in the context of atomic Bose-Einstein
condensates. This is, in particular, in the context of artificial
gauge fields more broadly, and more specifically spin-orbit
coupled Bose-Einstein condensates~\cite{dalibard}. There, admittedly,
the setup is somewhat different, as both the Dirac type operator
and the Schr{\"o}dinger one co-exist, but the compensating value
is that such settings have already been realized
experimentally~\cite{spiel2,spiel1,engels0,spiel3,weipan}. Moreover,
a wide range of coherent structures has been already proposed in
them including vortices~\cite{xuhan,spielpra},
Skyrmions \cite{skyrm}, Dirac monopoles \cite{dirac} and
dark solitons~\cite{brand,eplva},
as well as self-trapped states~\cite{santos}, bright solitons~\cite{prlva} and
gap-solitons~\cite{gapkono}.

In the present work, we will take a somewhat different path from
the above works. In particular, we will consider a relatively standard
example of the NLDE (namely the so-called massive Gross-Neveu model),
but motivated by the significant level of understanding and analytical
tractability afforded by discrete settings~\cite{pgk,pelinov}, we will
instead consider a {\it spatially discrete form} of the NLDE. A significant
part of our motivation for this consideration (and for the particular
form of the selected discretization) is due to (a) the possibility
to deploy the technology of the so-called anti-continuum (AC) limit
of MacKay-Aubry~\cite{macaub}, in order to appreciate the stability
properties near the limit of uncoupled adjacent sites and (b) the
feature that in the continuum limit of, in principle, infinite coupling,
our conclusions are expected to connect to what is known for the
corresponding PDE models that have been explored in the literature.
Admittedly, the discretization that is selected herein is, arguably,
not the most natural possible one (in that we utilize next-nearest
neighbors in order to discretize the first derivative terms by centered
differences). Nevertheless, it is identified that it is the most
suitable one for the present setting type of stencil and discrete solitary
waves are systematically obtained from the AC limit.
Moreover, a very recent development worth noting is that
spin-orbit Bose-Einstein condensates have recently been
considered in the realm of an optical lattice~\cite{engels},
which is often thought (in the so-called superfluid regime)
as being tantamount to a discretization of the original
continuum problem, through a suitable Wannier function
reduction~\cite{alfimov}. This suggests that considering
discrete variants of Dirac models may be a natural step for
near future considerations.

While many of our findings are somewhat reminiscent of the corresponding
discrete nonlinear Schr\"odinger (DNLS) equation ones~\cite{pgk,pelinov},
numerous others are rather unique to the Dirac equation. The single site
solution is found to lead to a rather unexpected waveform which we explain
and illustrate to effectively (that is, in its envelope) approach
in the continuum limit the solution of a different homoclinic
state problem that will be explicitly discussed below.
On the other hand, it is the two-site and three-site solutions
that lead to a continuation all the way to the continuum limit
of the Gross-Neveu solitary wave. However, contrary to what
is the case for the DNLS, the two-site solution turns out
to be stable close to the AC limit, while the three-site
solution is the unstable one close to that limit.
A count of the relevant eigenvalues near the AC limit is
systematically given for these different cases.
Subsequently
a near-alternation of stability is observed between these
two modes (the site- and inter-site-centered ones)
that is somewhat reminiscent of the phenomenology
identified in the saturable DNLS model~\cite{kip,champ}.
This is explored systematically, as is the feature of both
of these solutions in producing a complex quartet of modes
in a suitable band of the continuous spectrum. This oscillatory
instability and its dynamical by-products are traced as a function
of the propagation constant $\Lambda$ and of the inter-site
coupling strength $\epsilon$. The dynamical manifestation of the instabilities
within the discrete model is shown to lead to different possible features,
including the potential mobility of the solitary waves or their splitting
into multiple solitary waves of lower amplitude (and potentially of
a different type).

Our presentation is structured as follows. In section II, we present
an overview of our discrete model and its basic properties. In section III,
we examine the different solutions in the vicinity of the AC limit.
In section IV we examine the same solutions for large $\epsilon$, i.e.,
in the vicinity of the corresponding continuum limit. Finally, in section V,
we explore the dynamical instability manifestations of the different
solutions. Section VI summarizes our findings and presents our conclusions.

\section{Model and Theoretical Setup}

The NLDE model that we will consider  will be the massive Gross-Neveu
model with scalar-scalar interactions and a general
power-law nonlinearity. This is motivated by recent corresponding continuum
model explorations both at the level of mathematical analysis~\cite{comgus}
and at that of numerical computations~\cite{cooper,niur_recent}.
The discrete version of the equation introduced herein will be based
on a centered difference approximation of the first derivative in the form:

\begin{eqnarray}\label{eq:dyn}
i\dot U_n &=& \epsilon\nabla V_n - g (|U_n|^2-|V_n|^2)^k U_n + m U_n, \nonumber \\
i\dot V_n &=& -\epsilon\nabla U_n + g (|U_n|^2-|V_n|^2)^k V_n - m V_n ,
\end{eqnarray}
with $U_n$ and $V_n$ being the components of the spinor $\Psi_n\equiv(U_n,V_n)$ and $\nabla \Psi_n\equiv(\Psi_{n+1}-\Psi_{n-1})$ being the discrete gradient,
with a centered difference scheme, as indicated above. The
connection to the corresponding continuum limit can be assigned by
selecting $\epsilon=1/(2h)$ with $h$ being the lattice spacing (discretization parameter). It should also be noted in passing that we attempted to discretize
by a forward difference scheme, with considerably less promising results.
Given also that the centered difference scheme is a higher order discrete
approximation to the corresponding continuum limit, we therefore will only
present results by means of the centered difference discretization scheme
in what follows.

Our main focus hereafter will be on stationary solutions and their
stability.
Such solutions  can be found by using $U_n(t)=\exp(-i \Lambda t)u_n$, $V_n(t)=\exp(-i \Lambda t)v_n$, and satisfy the coupled algebraic equations:

\begin{eqnarray}\label{eq:stat}
\epsilon\nabla v_n - g (|u_n|^2-|v_n|^2)^k u_n + (m-\Lambda) u_n &=& 0, \nonumber \\
\epsilon\nabla u_n - g (|u_n|^2-|v_n|^2)^k v_n + (m+\Lambda) v_n &=& 0 .
\end{eqnarray}

Analogously to its continuum counterpart, the dynamical
system of Eq.~(\ref{eq:dyn})
presents a number of
 conserved quantities, such as the charge (squared $\ell^2$ norm):
\begin{equation}\label{eq:charge}
    Q=\sum_n \rho_n,\qquad \rho_n=|U_n|^2+|V_n|^2,
\end{equation}
with $\rho_n$ being the charge density, and the Hamiltonian:
\begin{equation}\label{eq:ham}
    H=\frac{1}{2}\sum_n \left[(U_n^*\nabla V_n-V_n^*\nabla U_n)-\frac{g}{k+1}(|U_n|^2-|V_n|^2)^{k+1}+m(|U_n|^2-|V_n|^2)\right].
\end{equation}

The dynamical equations (\ref{eq:dyn}) can be derived from the Hamiltonian (\ref{eq:ham}) by means of the Hamilton's equations:

\begin{equation}
    \mathrm{i}\dot U_n=\frac{\delta H}{\delta U_n^*},\qquad \mathrm{i}\dot{V_n}=\frac{\delta H}{\delta V_n^*} .
\end{equation}

Once stationary solutions of the algebraic system of Eqs.~(\ref{eq:stat}) are calculated (by e.g. fixed points methods), their linear stability is considered by means of a Bogoliubov-de Gennes linearized stability analysis.
More specifically, considering small perturbations [of order ${\rm O}(\delta)$, with $0< \delta \ll 1$] of the stationary solutions, we substitute the ansatz

\begin{equation}
    U_n(t)=e^{-i \Lambda t} \left[u_{n,0} + \delta (a_n e^{i \omega t} + c_n^{*} e^{-i \omega^{*} t}) \right],\ \ \
    V_n(t)=e^{-i \Lambda t} \left[v_{n,0} + \delta (b_n e^{i \omega t} + d_n^{*} e^{-i \omega^{*} t}) \right]
\end{equation}
into Eqs.~(\ref{eq:dyn}), and then solve the ensuing [to O$(\delta)$] eigenvalue problem:
\begin{equation}
    \omega
    \left(\begin{array}{c} a_n \\ b_n \\ c_n \\ d_n \end{array}\right)=\mathcal{M}
    \left(\begin{array}{c} a_n \\ b_n \\ c_n \\ d_n \end{array}\right),
\end{equation}
with $\mathcal{M}$ being
\begin{equation}\label{eq:stab}
    \mathcal{M}=\left(\begin{array}{cccc}
    \Lambda+J_n+L_n(u,u^*) & -\nabla-L_n(u,v^*) & L_n(u,u) & -L_n(u,v) \\
    \nabla-L_n(u^*,v) & \Lambda-J_n+L_n(v,v^*) & -L_n(u,v) & L_n(v,v) \\
    -L_n(u^*,u^*) & L_n(u^*,v^*) & \Lambda-J_n-L_n(u,u^*) & \nabla+L_n(u^*,v) \\
    L_n(u^*,v^*) & -L_n(v^*,v^*) & -\nabla+L_n(u,v^*) & \Lambda+J_n-L_n(v,v^*) \\
    \end{array}\right)
\end{equation}
for the eigenvalue $\omega$ and associated eigenvector $\{(a_n,b_n,c_n,d_n)^T\}$. Here, $L_n(x,y)$ is a function defined as:
\begin{equation}
    L_n(x,y)=k \chi_n^{k-1}x_{n,0}y_{n,0}\ ,
\end{equation}
with $J_n$ being:
\begin{equation}
    J_n\equiv g\chi_n^k-m\ ,
\end{equation}
and
\begin{equation}
    \chi_n\equiv |u_{n,0}|^2-|v_{n,0}|^2 .
\end{equation}

The dispersion relation of the linear excitations corresponds to the continuous spectrum that will be identified in the linearization around
the trivial $u_n=v_n=0\ \forall$  $n$ solution. This relation can be identified by decomposing the perturbations as
$\{a_n,b_n,c_n,d_n\}=\{A,B,C,D\}\exp(iqn)$ in Eqs.~(\ref{eq:stat}) and deriving the resulting condition:

\begin{equation}\label{eq:lineardisp}
    \omega(q)=\pm \Lambda\pm\sqrt{m^2+4\epsilon^2\sin^2q} .
\end{equation}

Consequently, there are two sets of bands in the essential spectrum. The embedded part given by $|\omega|\in[-\Lambda+m,-\Lambda+\sqrt{m^2+4\epsilon^2}]$, and
the non-embedded part, $|\omega|\in[\Lambda+m,\Lambda+\sqrt{m^2+4\epsilon^2}]$.
In what follows, for concreteness we will set $m=1$ and vary $\Lambda$,
as well as $\epsilon$, as our relevant parameters.

\section{Results from the AC limit (small coupling regime)} \label{sec:AC}

In this section, we consider the existence, stability and dynamics of discrete solitons from the AC to the continuum limit. In the AC limit, the soliton that can be continued up to the continuum is a three-site soliton, given by $v_n=0\ \forall$ $n$ and $u_{-1}=u_0=u_1=\left(1-\Lambda\right)^{1/(2k)}$, $u_n=0\ \forall$ $|n|\geq2$.

Let us explain below the general behavior for $\Lambda>1/3$. Outside
this range, the solitary waves are always unstable and hence we do not consider
them further here.

\begin{figure}
\begin{tabular}{cc}
\includegraphics[width=6cm]{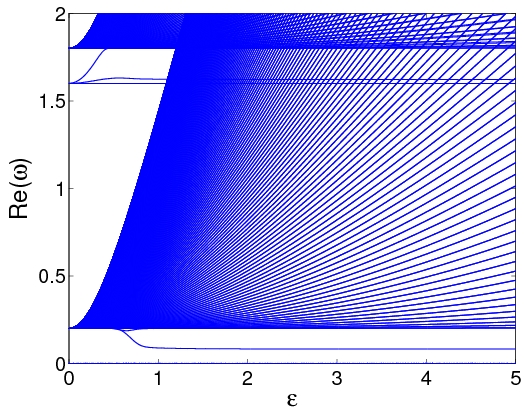} &
\includegraphics[width=6cm]{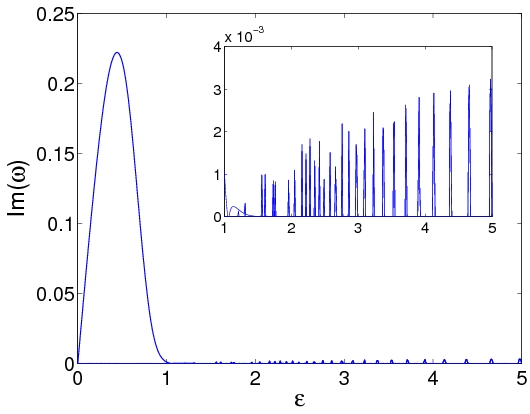} \\
\includegraphics[width=6cm]{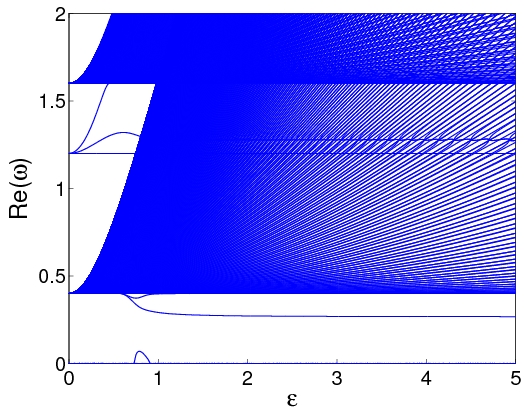} &
\includegraphics[width=6cm]{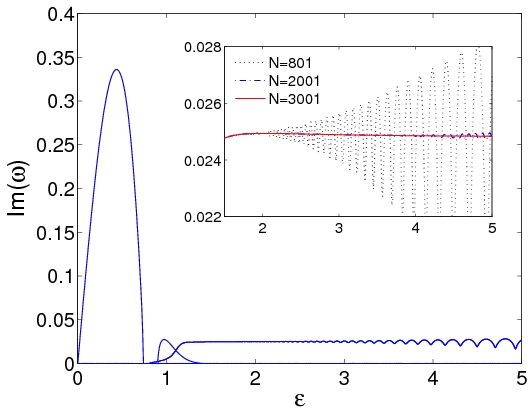}
\end{tabular}
\caption{Spectrum of the stability matrix (\ref{eq:stab}) for
discrete NLDE 3-site solitons with $\Lambda=0.8$ (top) and $\Lambda=0.6$ (bottom). The imaginary (left) and real (right) parts of the corresponding
eigenfrequencies are shown as a function of the coupling
strength $\epsilon$.
Only the positive real and imaginary parts of the eigenfrequencies are shown. The size of the system is $N=801$. In the top right, the inset is a magnification
of the relevant Im$(\omega)$ shown in the figure but at a different scale.
{The inset in the bottom right panel shows the oscillations of the growth rate for different system sizes when $\Lambda=0.6$. Notice that oscillation amplitude decreases rapidly as the number of lattice nodes increases.}}
\label{fig:stabdisc3s}
\end{figure}

It is easy to see from (\ref{eq:stab}) that in the AC limit and for any value of $k$, this three-site solution possesses 3 pairs of modes at $\omega=0$, 3 pairs at $\omega=\pm2\Lambda$, $(N-3)$ pairs at $\omega=\pm(1+\Lambda)$ and
$(N-3)$ pairs at $\omega=\pm(1-\Lambda)$. When the coupling is switched on (see Fig. \ref{fig:stabdisc3s}), the wave becomes exponentially unstable because of one among the 3 pairs at $\omega=0$ that detaches from the origin
yielding an imaginary eigenfrequency pair in a similar way as
occurs e.g. for the two-site structure in the DNLS equation~\cite{pgk}.
The other two vanishing eigenfrequency pairs remain at the origin. In addition, the eigenmodes at $\omega=\pm2\Lambda$ detach into three pairs that will subsequently collide with the embedded and essential parts of the spectrum; let us denote those modes as A, B, C (from upper to lower real part of the eigenfrequency). Mode C remains exactly at $\omega=\pm2\Lambda$ for every coupling. The real part of
the eigenfrequency of mode A rapidly increases entering the embedded spectrum at the point where the imaginary part of the eigenmode responsible for the exponential instability reaches its maximum. The exponential instability
mentioned previously
disappears close to (but not at) the point where mode C enters the essential spectrum. However, when the coupling increases, the exponential instability appears again with a similar (non-monotonic) behavior as the previous one, except for
the presence of smaller growth rates and of a slower decrease in the
growth rate (past the point of the maximal growth rate).
The most complex parametric dependence is the one experienced by mode B.
The latter enters the essential spectrum for a value of $\epsilon$ higher than
that for which mode C enters therein. Then, the system becomes oscillatorily
unstable and undergoes a Hopf bifurcation [in the case of finite
systems, due to the quantization of the continuous spectrum,
this translates
into a series of instability bubbles; for a similar scenario in the DNLS
see e.g.~\cite{johkiv}]. As a consequence, there are many oscillations in the imaginary part of mode B when the coupling is high; the amplitude of those oscillations decreases when the system size increases, as shown in the inset of bottom right panel of Fig. \ref{fig:stabdisc3s}]. When the frequency increases (say $\Lambda\gtrsim0.67$) the imaginary part of mode B does not asymptote to a nearly constant value
as the coupling strength increases, but, on the contrary, a series
of bubbles appears manifesting as oscillations around 0 (see top panels of Fig. \ref{fig:stabdisc3s}). The persistence of those oscillations in the continuum limit will be considered in Section \ref{sec:cont} (see also Fig. \ref{fig:bubbles} therein).

\begin{figure}
\begin{tabular}{cc}
\includegraphics[width=6cm]{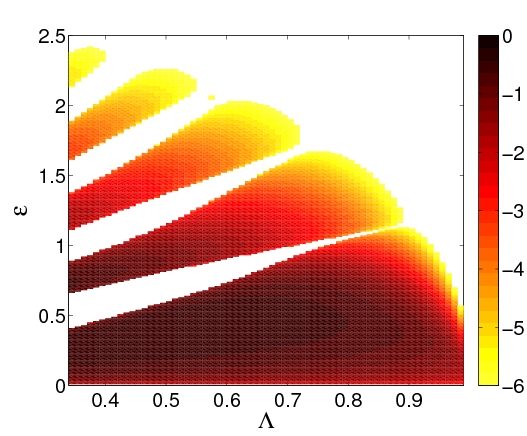} &
\includegraphics[width=6cm]{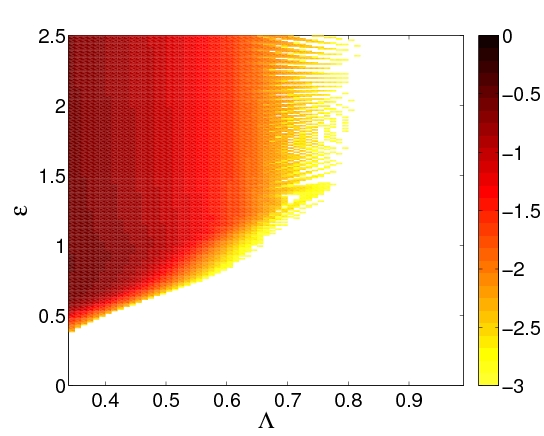}
\end{tabular}
\caption{Logarithm of the largest imaginary part of eigenvalues with zero (left) and non-zero (right) real part with $N=801$. Blank areas correspond to stable solitons.}
\label{fig:carpet}
\end{figure}

A more complete scenario of the linearization eigenfrequencies
is presented in Fig. \ref{fig:carpet}, where the largest imaginary part of
eigenfrequencies with zero and non-zero real part (i.e. responsible for exponential and Hopf bifurcations, respectively), with respect to $\epsilon\leq2.5$ and $1/3\leq\Lambda<1$ is presented. A cut-off for growth rates smaller than $10^{-3}$ for Hopf bifurcations and $10^{-6}$ for exponential bifurcations has been introduced. It can be observed that the instability bubbles emerge
in the right panel for $\Lambda\gtrsim0.67$ and $\epsilon\gtrsim1.08$. In
addition, it is observed that exponential instabilities emerge in several
lobes, which suggests a cascading mechanism of destabilizations and
restabilizations that we will return to below, upon examination of
the two-site solitary wave.

\begin{figure}
\begin{tabular}{cc}
\includegraphics[width=6cm]{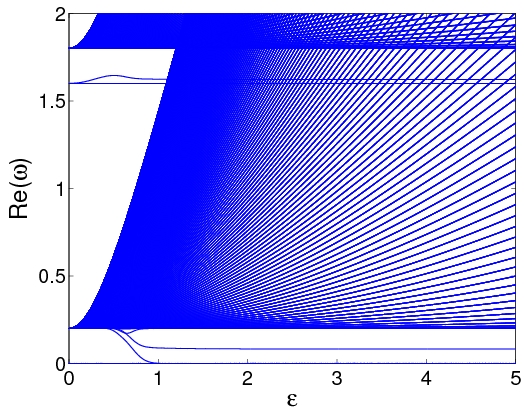} &
\includegraphics[width=6cm]{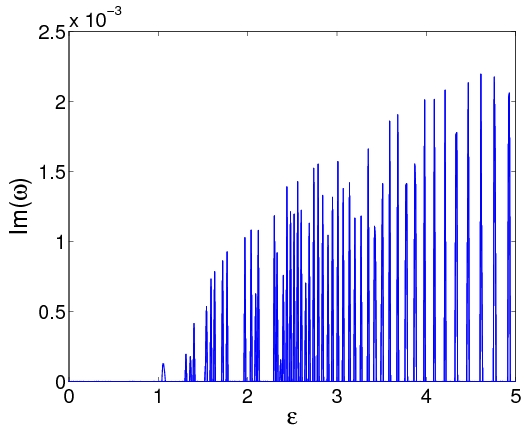} \\
\includegraphics[width=6cm]{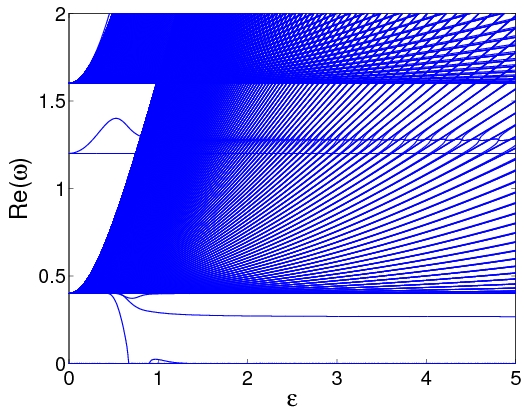} &
\includegraphics[width=6cm]{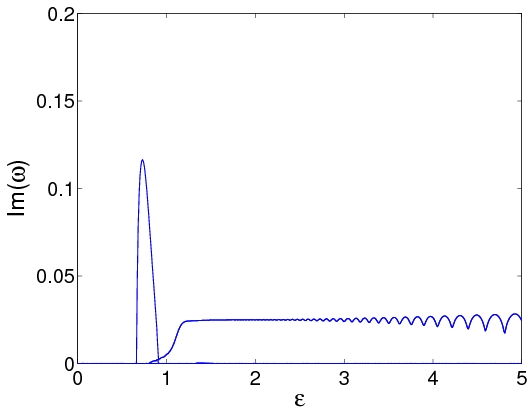}
\end{tabular}
\caption{Same as Fig.\ref{fig:stabdisc3s} but for the discrete
NLDE 2-site solitons with $\Lambda=0.8$ (top) and $\Lambda=0.6$ (bottom).}
\label{fig:stabdisc2s}
\end{figure}

\begin{figure}
\begin{center}
\includegraphics[width=6cm]{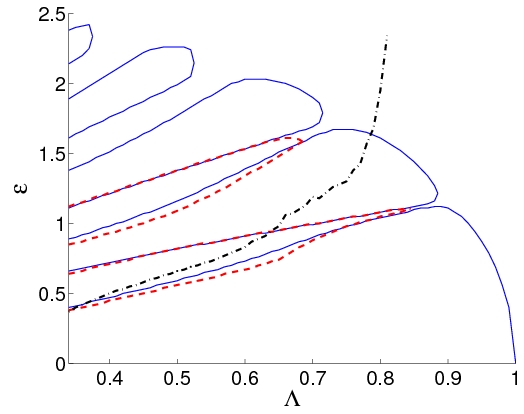}
\end{center}
\caption{$\epsilon$ vs $\Lambda$ plane where different unstable
regimes for 3-site and 2-site solitons are displayed. 3-site solitons are unstable inside the full line regions, whereas unstable 2-site solitons are inside dashed lines. Above the dashed-dotted lines, both solutions are oscillatorily unstable.
}
\label{fig:plane}
\end{figure}

A complementary scenario is experienced by the two-site solitary wave,
given in the AC limit by $v_n=0\ \forall$ $n$ and $u_0=u_1=\left(1-\Lambda\right)^{1/(2k)}$, $u_n=0$ elsewhere. At this limit, the 2-site structure
possesses 2 pairs of modes at $\omega=0$, 2 pairs at $\omega=\pm2\Lambda$, $(N-2)$ pairs at $\omega=\pm(1+\Lambda)$ and $(N-2)$ pairs at $\omega=\pm(1-\Lambda)$. When the coupling is switched on (see Fig. \ref{fig:stabdisc2s}), the
structure remains stable because of the persistence of both pairs at $\omega=0$. Mode A from $|\omega|=2\Lambda$ does not exist for this case;
on the other hand, the oscillatory instabilities caused by mode B also exist for the 2-site case. When increasing the coupling, the soliton experiences an exponential bifurcation and becomes unstable, contrary to the 3-site soliton (notice that in typical Klein-Gordon and --e.g. saturable-- DNLS settings,
such stability exchanges take place between 2-site and 1-site breathers or solitons). Here, there are exponential stability exchanges between 2-site and 3-site solitons, although the bifurcations of the two families of solutions
do not perfectly coincide (nevertheless, in a number of such
exchanges, the corresponding stabilization/destabilization thresholds
are fairly proximal). This scenario is summarized in Fig. \ref{fig:plane}.
We should note in passing that these near-exchanges of stability
suggest a scenario similar to the ones occurring e.g. in the
saturable or cubic-quintic DNLS model where the near-exchange of
stability of the 1- and 2-site solitary waves (in that case) is mediated
through a series of pitchfork and reverse pitchfork bifurcations
of asymmetric solution branches~\cite{cqdnls,johrod}. However, we will
not pursue the relevant narrow branches of asymmetric solutions
herein.

\begin{figure}
\begin{tabular}{cc}
\includegraphics[width=6cm]{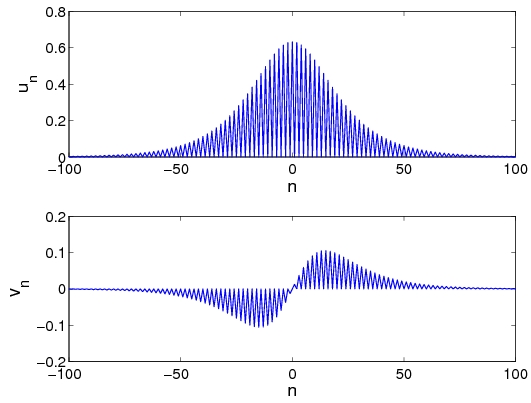} &
\includegraphics[width=6cm]{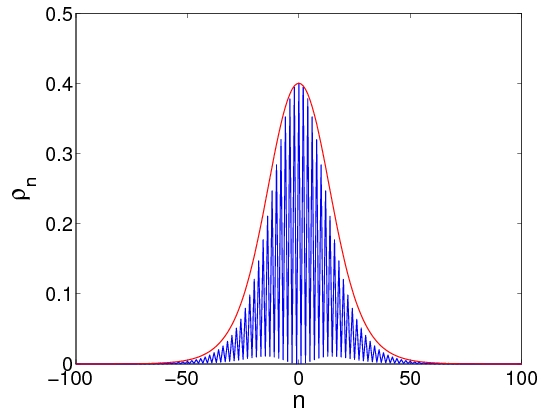} \\
\end{tabular}
\caption{(Left) The two-component profiles
for a 1-site soliton with $\Lambda=0.8$ at $\epsilon=5$. (Right) Charge density $\rho_n=|u_n|^2+|v_n|^2$ for the solitary wave at the left (blue line) and a 3-site soliton with the same parameters (red line). It is clear that the former
does {\it not} asymptote to the latter, but rather to a different envelope
that will be revealed in section IV below.}
\label{fig:prof1s}
\end{figure}

There is an interesting kind of solution that also
exists from the AC limit and can be continued to the continuum limit, namely the one-site soliton. This has the following property which is, in fact,
preserved upon continuation for any value of the coupling (see Fig. \ref{fig:prof1s}): $u_n=0$ for odd $n$ and $v_n=0$ for even $n$; however, the charge density of the soliton is qualitatively different from that of the three-site solitons. In the AC limit $\epsilon=0$, $u_0=\left(1-\Lambda\right)^{1/(2k)}$, and $u_n=0$ for the rest of sites (with $v_n=0\ \forall$ $n$).

The form of this solution can be identified as we approach the continuum limit as $u_n=0$ for odd $n$ and $v_n=0$ for even $n$, by transforming the discrete NLDE equation (\ref{eq:dyn}) into the new set of equations:

\begin{eqnarray}\label{eq:stat1s}
\epsilon(v_{n+1}-v_{n-1}) - g u_n^{2k+1} + (m-\Lambda) u_n &=& 0,\ \mathrm{for~even }~n  \nonumber \\
\epsilon(u_{n+2}-u_n) - (-1)^{k} g v_{n+1}^{2k+1} + (m+\Lambda) v_{n+1} &=& 0,
\ \mathrm{for~even }~n
\end{eqnarray}
which possesses homoclinic solutions in the continuum limit (see Section \ref{sec:cont}).

\begin{figure}
\begin{tabular}{cc}
\includegraphics[width=6cm]{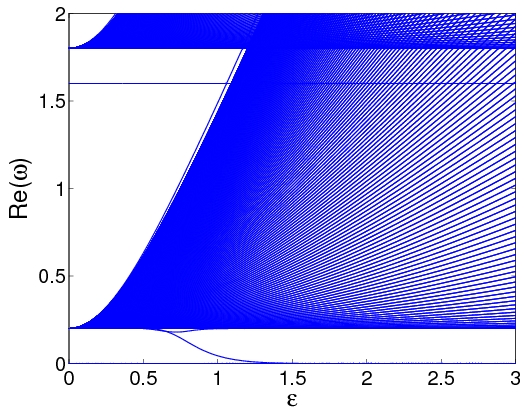} &
\includegraphics[width=6cm]{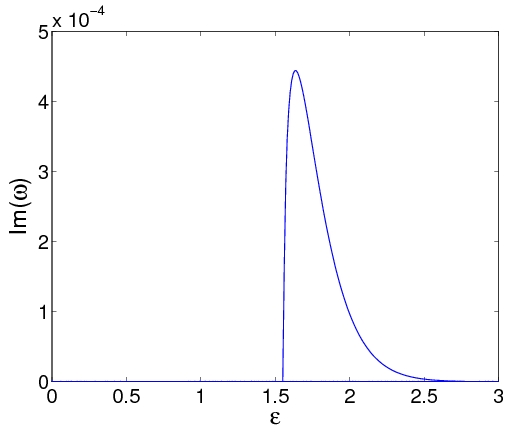} \\
\includegraphics[width=6cm]{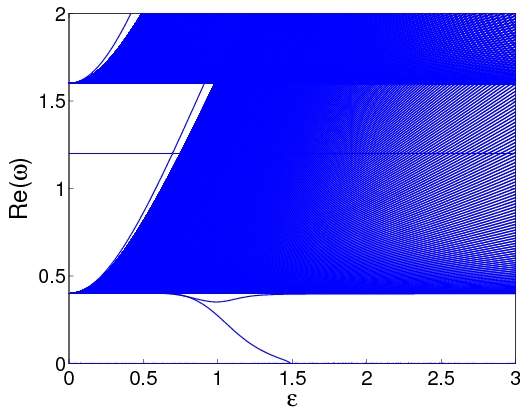} &
\includegraphics[width=6cm]{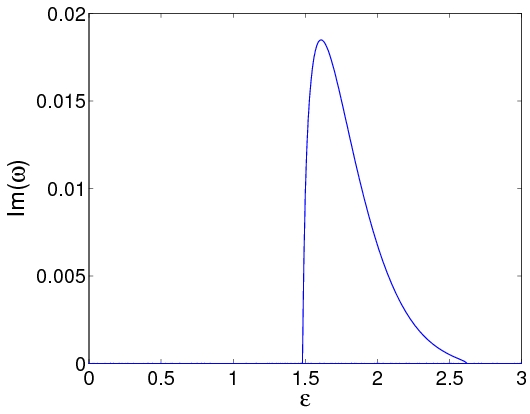}
\end{tabular}
\caption{Same as Fig.\ref{fig:stabdisc3s} but for discrete NLDE 1-site solitons with $\Lambda=0.8$ (top) and $\Lambda=0.6$ (bottom).}
\label{fig:stabdisc1s}
\end{figure}

\begin{figure}
\begin{tabular}{cc}
\includegraphics[width=6cm]{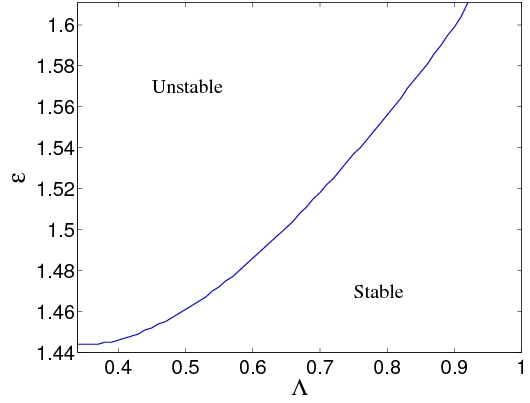} &
\includegraphics[width=6cm]{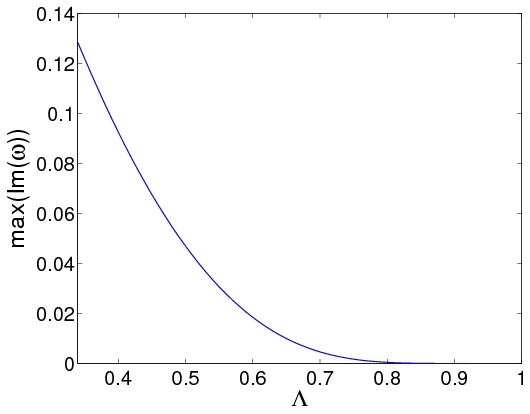} \\
\end{tabular}
\caption{Exponential instability loci (left) for a 1-site soliton
are shown in the $\epsilon$-$\Lambda$ plane in the left panel. The right
panel shows the maximum (over the considered $\epsilon$ variations)
growth rate at each value of $\Lambda$.
For $\Lambda>0.92$, the growth rates for the 1-site soliton are smaller
than $10^{-7}$ and cannot be accurately traced because of machine precision.
}
\label{fig:plane1s}
\end{figure}

The spectrum of the one-site solitons at $\epsilon=0$ consists of a single pair of eigenvalues at $\omega=0$ and another single pair at $\omega=\pm2\Lambda$; apart from these, there are $N-1$ pairs at $\omega=\pm(1+\Lambda)$ and $\omega=\pm(1-\Lambda)$. When the coupling is switched on, as there is only a single pair of eigenmodes at $\omega=0$, the soliton does not experience exponential bifurcations; in addition, the non-existence of mode B prevents the existence of harmful Hopf bifurcations arising in 3-site and 2-site solitons (see Fig. \ref{fig:stabdisc1s}). The only observed instability is an exponential one arising for a finite value of coupling and caused by a mode that bifurcates
from  the essential spectrum as the coupling strength increases;
the growth rate which has a non-monotonic dependence on
the coupling and tends asymptotically to zero when reaching the continuum limit, and its maximum value decreasing with $\Lambda$ (for fixed coupling)
are shown in Fig. \ref{fig:plane1s} for more details. Similar to the 3-site
structures, there is a complementary family of solitons consisting of
2-site structures with a hole in between, characterized by $u_0=u_2$.

\section{Results near the continuum
limit (large coupling regime)} \label{sec:cont}

In this section, on the one hand, we will connect
the findings of our model with some previous
results about the stability of the continuous NLDE. On the other hand, we
will perform the Bogoliubov-de Gennes (BdG) spectral stability analysis
of the discrete NLDE for a large coupling such as $\epsilon=5$, which
corresponds to
a spatial discretization parameter $h=0.1$.

Previous results from Comech (see Refs. \cite{comech1,comech2,comech3,boucom}) show that close to the non-relativistic limit ($\Lambda\lesssim 1$), the Vakhitov--Kolokolov criterion should hold. Based on it \cite{comech2}, it is concluded that no unstable eigefrequency should emerge from $\omega=0$ close to this
limit for $k=1$ or $k=2$ (contrary to the $k\geq3$, $k\in\mathbb{N}$ case
where a pair of eigenfrequencies with a nonzero imaginary part and a zero
real part are present) and, consequently, no exponential instability
should exist in that limit. Additionally, for any $k$, the
existence of an eigenfrequency $|\omega|=2\Lambda$ is also predicted.
This mode enters the linear mode band at
$\Lambda=1/3$ (when increasing $\Lambda$).

The work of~\cite{niurka}, based on the so-called Bogolubsky criterion,
as well as that of~\cite{boucom} suggest that solitary waves
are always unstable for $\Lambda<\Lambda_c$. [It is
worth noting here that neither of the two criteria mentioned above
is able to give a necessary condition and, consequently, the minimum value for which solitons are stable must be determined numerically]. In the cubic case
($k=1$) it is predicted in~\cite{niurka}
that $\Lambda_c=0.6976$. However, in a recent paper \cite{niur_recent}, further
numerical simulations have suggested that solitons may be dynamically stable
for $\Lambda\geq0.56$.

\begin{figure}
\begin{tabular}{cc}
\includegraphics[width=6cm]{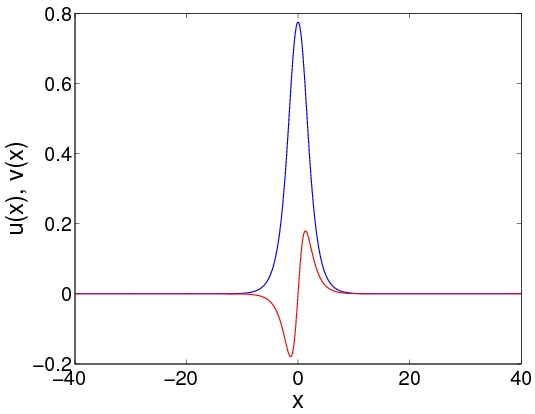} &
\includegraphics[width=6cm]{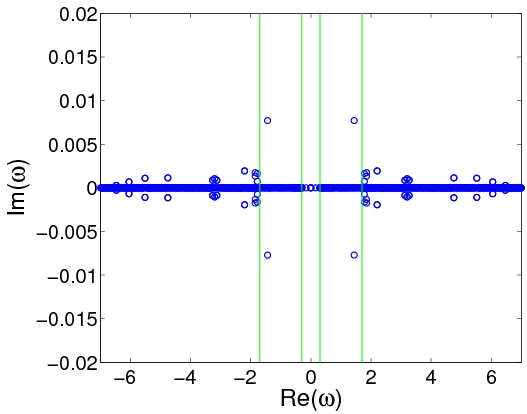} \\
\includegraphics[width=6cm]{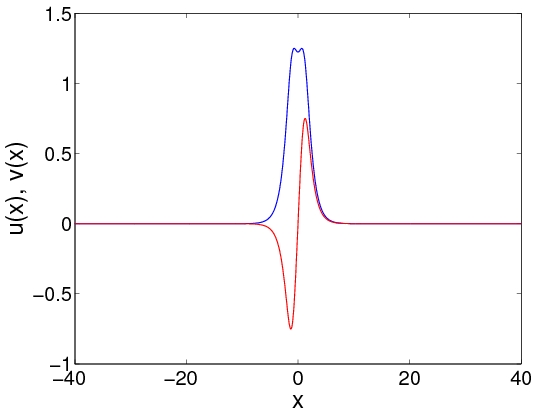} &
\includegraphics[width=6cm]{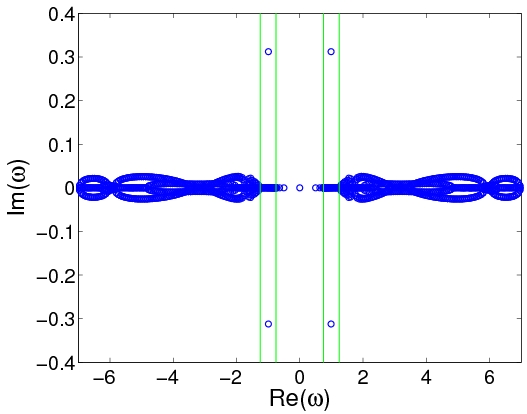} \\
\end{tabular}
\caption{(Left) Solitary wave profiles and (right) spectral planes
for $h=0.1$, $L=80$ and $\Lambda=0.7$ (top) and $\Lambda=0.25$ (bottom). Lines in the right figures indicate the limits of the embedded and non-embedded parts of the essential spectrum. Instabilities in the non-embedded spectrum should vanish when $L\rightarrow\infty$.}
\label{fig:profcont}
\end{figure}

The analytical form of the profile of solitons in the continuum limit is given by \cite{cooper,niurka}:

\begin{equation}
    u(x)=\sqrt{\frac{(1+\Lambda)\cosh^2(k\beta x)}{1+\Lambda\cosh(2k\beta x)}}\left[\frac{(k+1)\beta^2}{1+\Lambda\cosh(2k\beta x)}\right]^{1/2k},\quad
    v(x)=\sqrt{\frac{(1-\Lambda)\sinh^2(k\beta x)}{1+\Lambda\cosh(2k\beta x)}}\left[\frac{(k+1)\beta^2}{1+\Lambda\cosh(2k\beta x)}\right]^{1/2k} ,
\end{equation}
with $\beta=\sqrt{1-\Lambda^2}$. When $k=1$ the equations above can be simplified to:

\begin{equation}
    u(x)=\frac{\sqrt{2(1-\Lambda)}}{[1-\mu\tanh^2(\beta x)]\cosh(\beta x)},\quad
    v(x)=\frac{\sqrt{2\mu(1-\Lambda)}\tanh(\beta x)}{[1-\mu\tanh^2(\beta x)]\cosh(\beta x)} ,
\end{equation}
with $\mu=(1-\Lambda)/(1+\Lambda)$. As demonstrated in \cite{cooper}, continuous solitons become double-humped for $\Lambda$ smaller than a critical value for every $k$. Fig. \ref{fig:profcont} shows the profile and spectral planes for two different examples of solitons close to the continuum limit with $k=1$.

\begin{figure}
\begin{tabular}{cc}
\includegraphics[width=6cm]{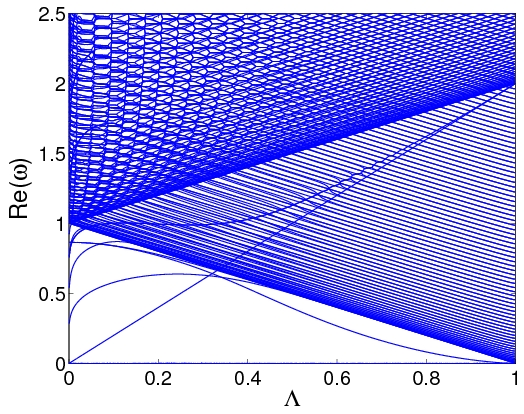} &
\includegraphics[width=6cm]{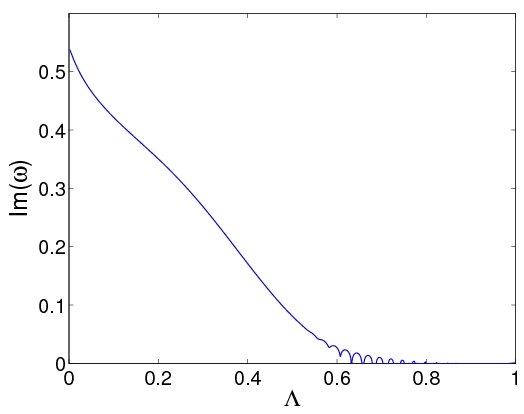}
\end{tabular}
\caption{Spectrum of the stability matrix (\ref{eq:stab}) for solitons with $h=0.1$ and $L=80$. For the sake of simplicity, only the positive real and imaginary parts of the eigenvalues are shown.}
\label{fig:stabcont}
\end{figure}

We show in Fig. \ref{fig:stabcont} the stability eigenvalues for $k=1$ in a domain $x\in[-L/2,L/2]$, with $L=80$ and a discretization step $h=0.1$. Although there are instabilities caused by eigenvalue collisions in the non-embedded spectrum, we have neglected them, as they disappear in the limit of $h\rightarrow0$ and $L\rightarrow\infty$. The waves are found to be unstable for small $\Lambda$,
with a growth rate that decreases when $\Lambda$ is increased. The source of instabilities is a localized mode (with non-zero imaginary part of
its eigenfrequency even when $\Lambda\rightarrow0$) that enters the embedded spectrum at $\Lambda\approx0.037$. Once inside the linear modes band, this
localized mode causes multiple bubbles, yet
at $\Lambda\approx0.632$, it returns to the real eigenfrequency
axis and the solitary wave becomes stable. Nevertheless,
this stability is ephemeral, as the soliton becomes unstable again at $\Lambda\approx0.634$. From this point, there is a succession of instability bubbles, whose amplitude (i.e., the maximal growth rate associated with them)
decreases with $\Lambda$. Notice also the existence of the eigenvalue with $\omega=2\Lambda$, which enters the embedded spectrum at $\Lambda=1/3$.

\begin{figure}
\begin{tabular}{cc}
\includegraphics[width=6cm]{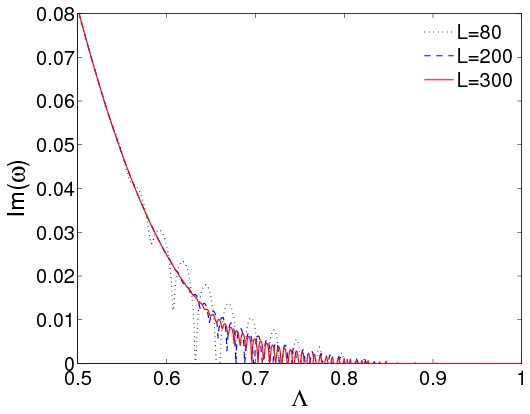}
\end{tabular}
\caption{Dependence of the growth rates of the solitary waves
shown in Fig. \ref{fig:stabcont} for different domain lengths.
As $L$ increases, we progressively can discern the envelope of the
infinite domain limit.}
\label{fig:bubbles}
\end{figure}

In order to observe the behavior of bubbles when the domain is enlarged, we have included Fig. \ref{fig:bubbles} where the growth rate is plotted for $L=80$, $200$ and $300$. It is observed that the number of bubbles increases with $L$, but their width decreases. In any case, the envelope of the bubbles tends to zero asymptotically when $\Lambda$ approaches 1, in a similar way as it was observed for dark solitons in DNLS  settings~\cite{johkiv}.
Unfortunately, the convex nature of the relevant (apparent) envelope
curve is inconclusive in connection to the stability aspect.  In
particular, it is unclear, based on the present computations,
whether the curve, as $h \rightarrow 0$, {\it still} intersects
the axis and no longer features an unstable mode past a
critical $\Lambda_c$, as is the case
with our finite $h$, finite domain computations in Fig.~\ref{fig:bubbles},
The alternative scenario is that the approach to the stable
NLS limit of $\Lambda \rightarrow 1$ (a glimpse of
which is illustrated in Fig.~\ref{fig:bubbles}) is merely asymptotic.
It would be especially interesting to pursue this intriguing aspect
further, pushing the envelope of the currently available numerical
tools.

As mentioned in Section \ref{sec:AC}, the 1-site solitons can also exist in the continuum limit. There, by neglecting the irrelevant (in this setting)
inactive odd sites for one of the fields, and the even ones
for the other, the envelope of the solitary waves can be seen
to approach the homoclinic orbits of the following system of ODEs that is found by obtaining the continuum limit of (\ref{eq:stat1s}):

\begin{eqnarray}\label{eq:homoc}
\partial_x u &=& (-1)^k g v^{2k+1} - (m+\Lambda) v , \nonumber \\
\partial_x v &=& g u^{2k+1} - (m-\Lambda) u_0  .
\end{eqnarray}
Using phase plane numerical analysis (not shown here), we have
confirmed that Eqs.~(\ref{eq:homoc}) possess a homoclinic orbit
$g=m=1$ and $k=1$, for a wide range of $\Lambda$'s. We have
also confirmed that it is at these very homoclinic orbits
that the envelope of our NLDE 1-site solution tends as
the coupling strength is increased.

\begin{figure}
\begin{tabular}{cc}
\includegraphics[width=6cm]{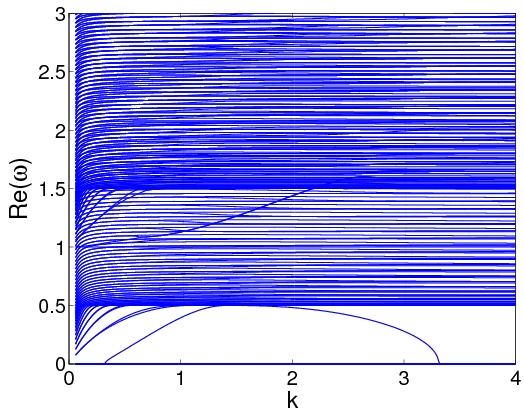} &
\includegraphics[width=6cm]{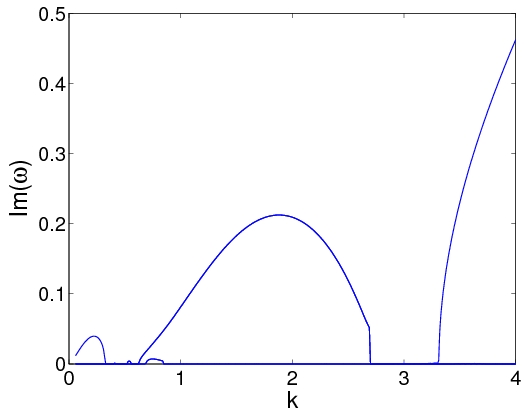} \\
\includegraphics[width=6cm]{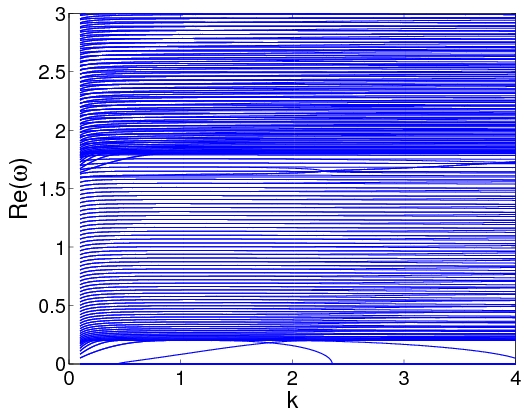} &
\includegraphics[width=6cm]{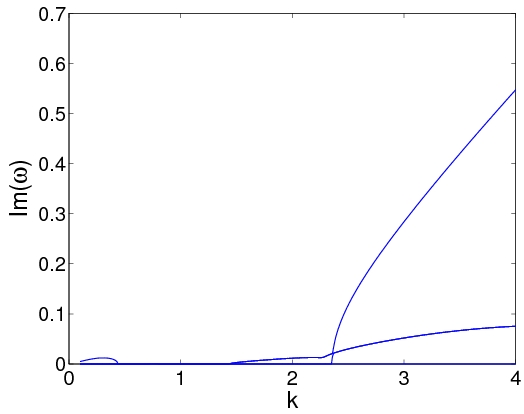} \\
\end{tabular}
\caption{Spectrum of the stability matrix (\ref{eq:stab}) for solitons with $h=0.1$ and $L=80$. $\Lambda=0.5$ in upper panels and $\Lambda=0.8$ in bottom ones, in this case as a function of the nonlinearity exponent $k$. In the former case, exponential instabilities emerge for $k<0.33$ and $k>3.35$, whereas in the latter case, those instabilities take place for $k<0.44$ and $k>2.36$.}
\label{fig:stabkvar}
\end{figure}

As a final comment regarding the stability analysis, we note
that our approach allows to examine not only the variations
as a function of the propagation constant $\Lambda$, as well as
the coupling strength $\epsilon$, but additionally also with
respect to the nonlinearity exponent parameter $k$.
Fig. \ref{fig:stabkvar} shows some typical
examples of this variation for $h=0.1$ and
values of $\Lambda=0.5$ (top) and $\Lambda=0.8$ (bottom).
The parametric variation of $k$ reveals
both the Hopf and exponential instabilities of the system.
As regards the latter, we note that for sufficiently high values of
$k$, an eigenfrequency bifurcating from the continuous spectrum
crosses the spectral plane origin becoming imaginary, in accordance
with the expectation that for sufficiently high $k$ a blow-up type
instability (which for $\Lambda \rightarrow 1$, i.e., the NLSE
limit, should occur for $k > 2$) should emerge.  The relevant
critical points are $k=3.35$ and $k=2.36$, respectively for
the considered values of $\Lambda$ of $0.5$ and $0.8$.
On the other hand, another
interesting observation is that a similar bifurcation to
exponential instability appears to emerge in the small
positive $k$ i.e., the weakly nonlinear limit. This instability
arising for $k <0.33$ and $k<0.44$ in the top and bottom, respectively,
panel of Fig. \ref{fig:stabkvar} is worth examining further
in its own right, possibly
through a perturbative calculation.

\begin{figure}
\begin{center}
\includegraphics[width=6cm]{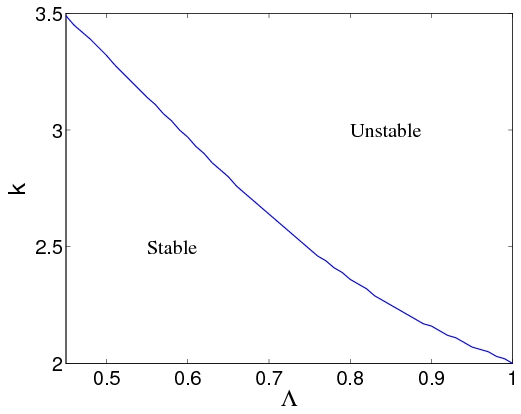}
\end{center}
\caption{$k$ vs $\Lambda$ plane where the behavior of solitons with respect to exponential instabilities is displayed. Notice that the critical value $k=2$ is
retrieved at the non-relativistic (NLSE) limit.}
\label{fig:kvarplane}
\end{figure}

A systematic exploration of the $k$-$\Lambda$ plane of the relevant
exponential instability is identified in Fig.~\ref{fig:kvarplane}.
As expected in the non-relativistic NLSE limit of $\Lambda \rightarrow
1$,
no exponential instabilities are observed when $k<2$, whereas for $k>2$, the solitons are unstable i.e., amenable to collapse.
 We can see that as $\Lambda$ decreases from that limit,
the corresponding critical $k$ for the instability monotonically
increases.

\section{Dynamical Evolution of Instabilities}

We have also briefly analyzed the dynamics of unstable 3-site solutions in different regimes. Below, we give a number of selected results in connection to
the relevant numerical evolution,
although admittedly a systematic classification
of the dynamical implications of the different identified instabilities
and of the various possible configurations identified herein is a separate
numerical project in its own right.

\begin{figure}
\begin{tabular}{cc}
\includegraphics[width=6cm]{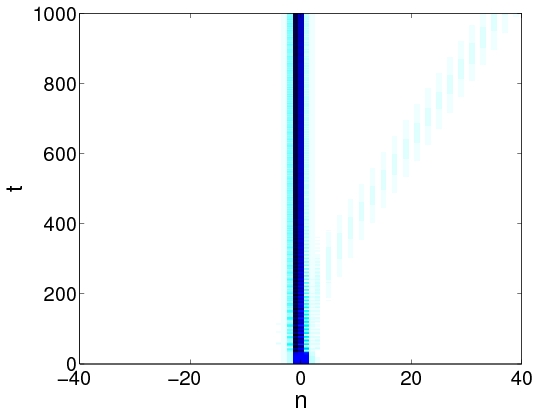} &
\includegraphics[width=6cm]{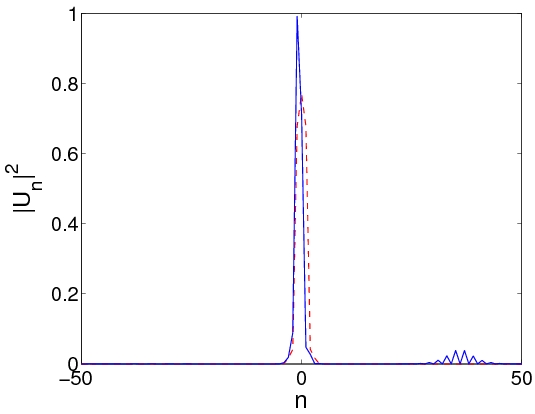} \\
\includegraphics[width=6cm]{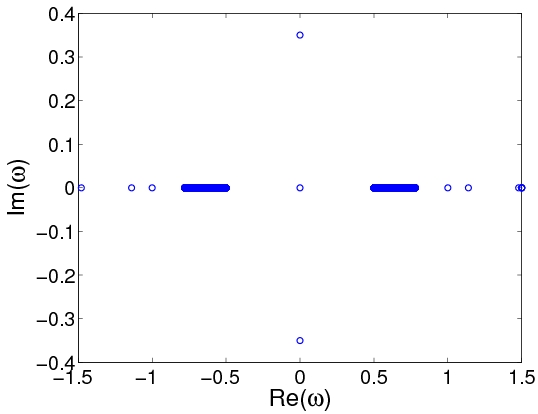} &
\includegraphics[width=6cm]{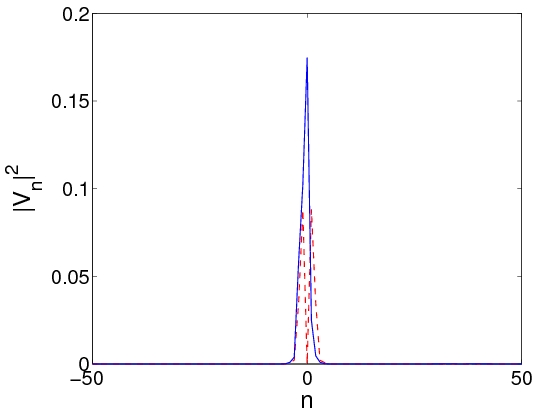} \\
\end{tabular}
\caption{Evolution of the soliton with $\Lambda=0.5$ and $\epsilon=0.4$. Top panel shows the evolution of the charge density $\rho_n$. Right panels depict the fields at $t=0$ (dashed line) and at $t=1000$ (solid line). Bottom left panel displays the spectral plane of the solitary wave.}
\label{fig:simul1}
\end{figure}

\begin{figure}
\begin{tabular}{cc}
\includegraphics[width=6cm]{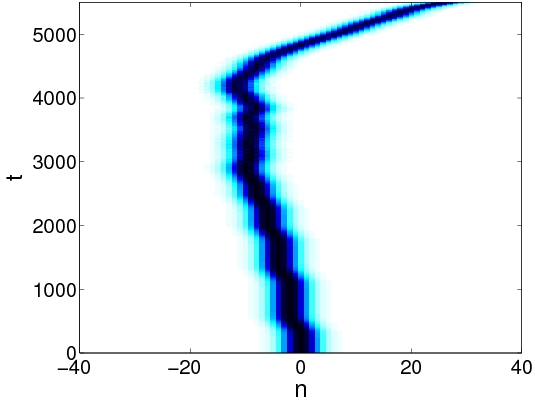} &
\includegraphics[width=6cm]{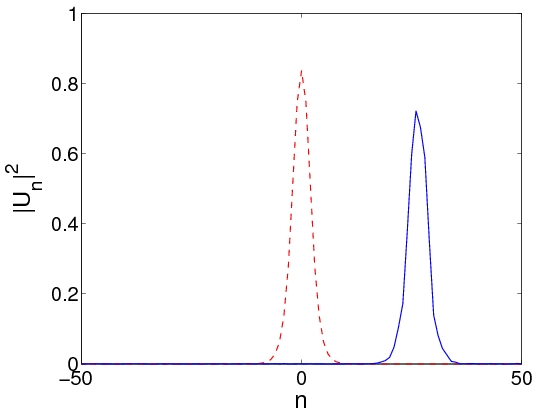} \\
\includegraphics[width=6cm]{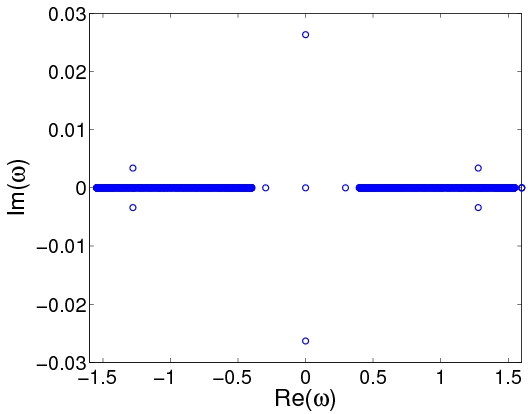} &
\includegraphics[width=6cm]{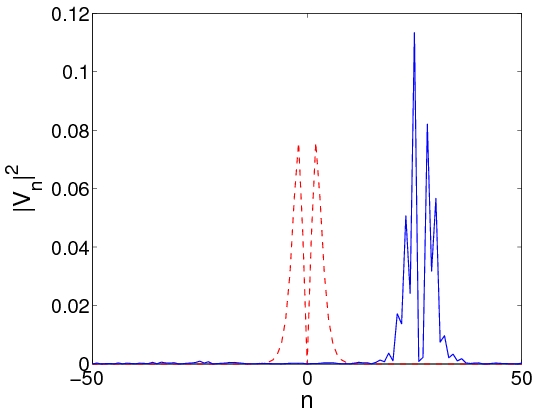} \\
\end{tabular}
\caption{Evolution of the soliton with $\Lambda=0.6$ and $\epsilon=0.95$. Top panel shows the evolution of the charge density $\rho_n$. Right panels depict the fields at $t=0$ (dashed line) and at $t=5500$ (solid line). Bottom left panel displays the spectral plane of the solitary wave.}
\label{fig:simul2}
\end{figure}

Fig. \ref{fig:simul1} shows the evolution of a solitary wave
with $\Lambda=0.5$ and $\epsilon=0.4$, i.e. inside the lower lobe of exponential instabilities. We observe that the structure emits linear
wave ``radiation'' and subsequently deforms towards
a more compact configuration with fewer high-amplitude excited
sites (more specifically one in each component). If a solution within the intermediate lobe is taken (as e.g. that of Fig. \ref{fig:simul2}, where $\Lambda=0.6$ and $\epsilon=0.95$), it is observed that the soliton moves along the lattice. In this regime, both an exponential
and an oscillatory instability are present. Generally, for cases
of larger coupling, we find that the solutions are more
prone to becoming mobile, upon the manifestation of the dynamical
instability.

\begin{figure}
\begin{tabular}{cc}
\includegraphics[width=6cm]{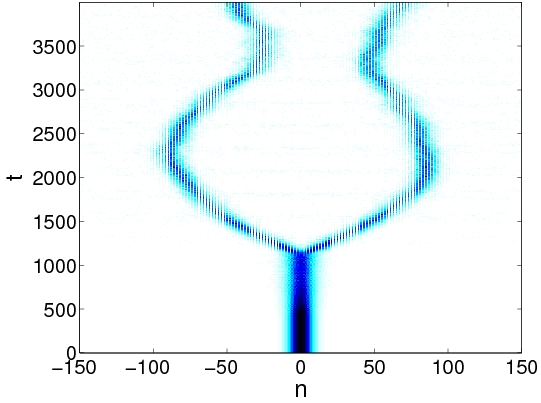} &
\includegraphics[width=6cm]{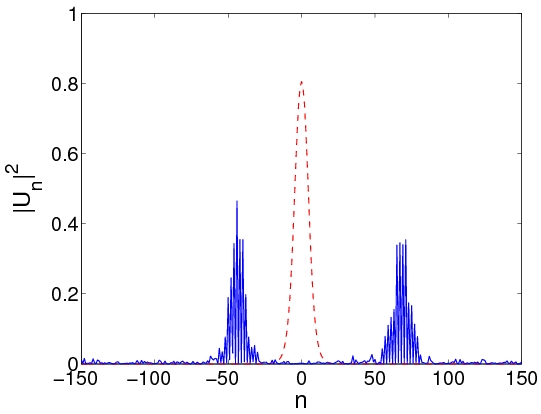} \\
\includegraphics[width=6cm]{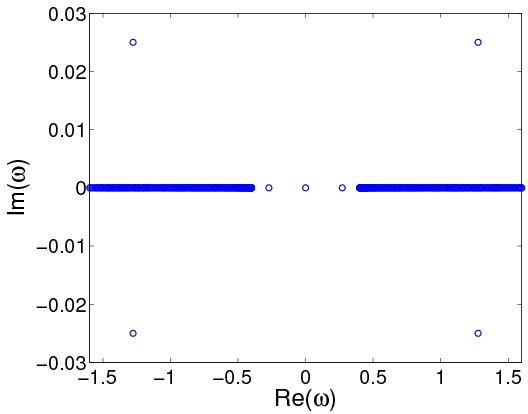} &
\includegraphics[width=6cm]{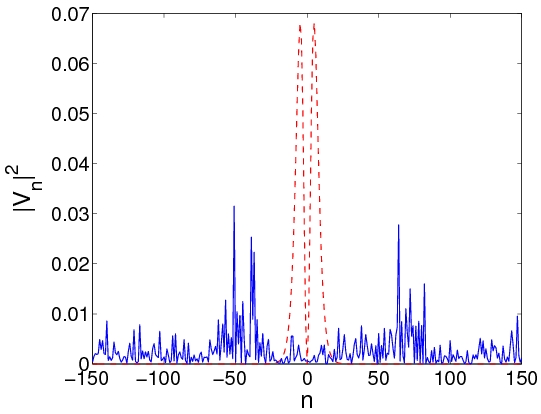} \\
\end{tabular}
\caption{Evolution of the soliton with $\Lambda=0.6$ and $\epsilon=2$. Top panel shows the evolution of the charge density $\rho_n$. Right panels depict the fields at $t=0$ (dashed line) and at $t=4000$ (solid line). Bottom left panel displays the spectral plane of the solitary wave.}
\label{fig:simul3}
\end{figure}

\begin{figure}
\begin{tabular}{cc}
\includegraphics[width=6cm]{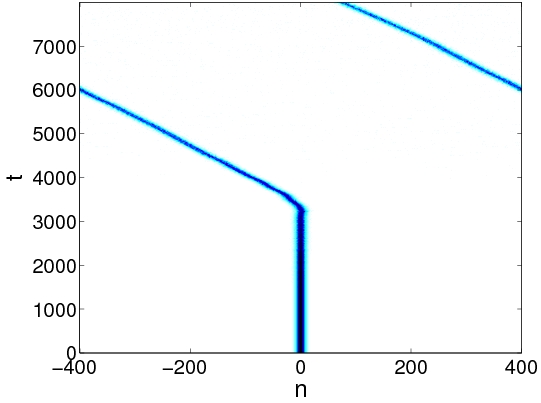} &
\includegraphics[width=6cm]{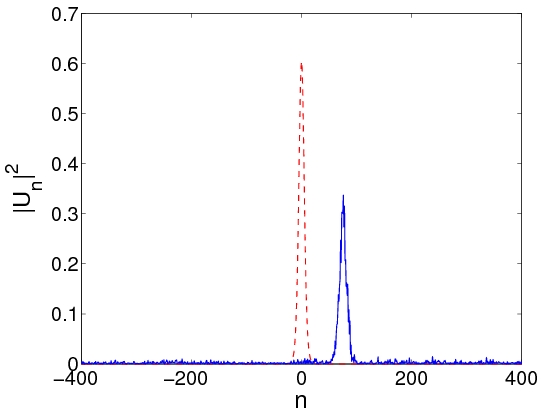} \\
\includegraphics[width=6cm]{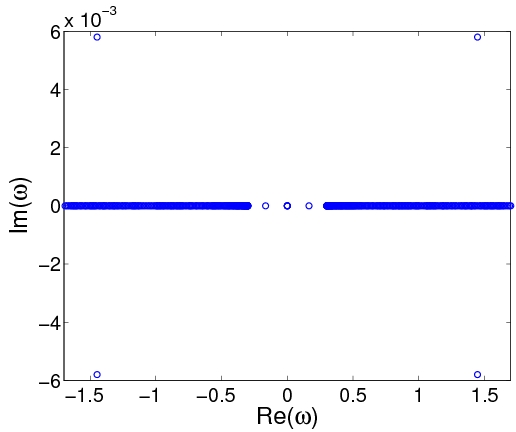} &
\includegraphics[width=6cm]{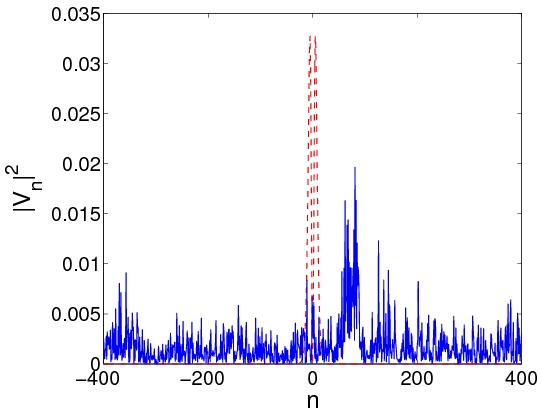} \\
\end{tabular}
\caption{Evolution of the soliton with $\Lambda=0.7$ and $\epsilon=2$. Top panel shows the evolution of the charge density $\rho_n$. Right panels depict the fields at $t=0$ (dashed line) and at $t=8000$ (solid line). Bottom left panel displays the spectral plane of the solitary wave.}
\label{fig:simul4}
\end{figure}

Fig. \ref{fig:simul3} depicts an oscillatorily unstable wave with $\Lambda=0.6$ and $\epsilon=2$. Interestingly, the latter splits into two
daughter-waves as a result of the oscillatory growth. Once the original
structure is split, the charge density at even sites is close to zero (a state similar to the 1-site soliton).
That is, the offspring in this case belong to the same class of
solitary waves as the 1-site solution examined above.
If an oscillatory unstable soliton is taken within the region of oscillatory
instability bubbles (see Fig. \ref{fig:simul4}, where $\Lambda=0.7$ and $\epsilon=2$), the soliton is put into motion. Once again, this is a relatively
common feature of case examples with large values of $\epsilon$ that
are more proximal to the continuum limit of the problem. Notice, however,
additionally that in the process of shedding away radiative wavepackets
that manifests the dynamical instability and sets the solitary wave
in motion, the amplitude of the structure decreases, which indicates
that its effective $\Lambda$ increases and hence renders it more
robust.



Lastly, we should note that we have also examined dynamical
instabilities of other structures such as 1- and 2-site solitons.
These often, too, result in mobile coherent structures, especially
for large values of $\epsilon$, although some states, such as
the staggered 1-site wave are less amenable to extensive traveling
throughout the lattice, perhaps partly due to their special
spatial structure.



\section{Conclusions and Future Challenges}

In the present work we have examined a lattice analogue of the
nonlinear Dirac equation. Motivated by the considerable volume
of both mathematical and computational investigations of the
continuum limit of the corresponding problem, we have developed
a prototypical discretization scheme whose continuum limit
is the Gross-Neveu model. This is a model that while it does not
presently possess a straightforward physical realization (e.g.
analogous to what is the case for its DNLS analogue
and, say, optical waveguide arrays~\cite{segev}), it nevertheless is
of substantial interest in its own right for a number of
reasons. It is useful (and relevant to understand),
on the one hand, as a numerical
scheme and a computational tool
for approximating the corresponding continuum limit
(in regimes of large coupling strength $\epsilon$). On the other hand,
its analytical tractability in the vicinity of the anti-continuum
limit of uncoupled sites makes it a useful starting point for
the exploration of the spectral properties of solitary waves.
In the AC limit, there is a complete control over these spectral
properties and corresponding eigenvalues, and it then remains
to appreciate the continuation of these over the coupling strength
$\epsilon$, in order to understand both the features of the
discrete model and those of its continuum limit. Moreover,
the physical realization of quasi-discrete systems possessing
Dirac-like dynamics such as spin-orbit Bose-Einstein condensates
in the presence of optical lattices very recently~\cite{engels},
seems to strongly suggest the potential experimentally-relevant realization
of models of this class in the near future.

In light of the above motivations, here
we have shown a multitude of unexpected properties
that merit further studies not only from a computational but also
importantly from a rigorously mathematical point of view.
In particular, we showed that a single site excitation does not
continue, as might be expected, to a continuum solitary wave of
the Gross-Neveu model. Instead, it forms a remarkable staggered
structure that approaches in the limit of $\epsilon$ large
(while being preserved as a state) the envelope of
the homoclinic orbit of a different dynamical model.
This appears to be a fairly robust structure
in its parametric dependence over $\epsilon$ and $\Lambda$.
On the other hand, the two- and three-site
initial excitations play the role, respectively, of the one- and two-site
excitations of the DNLS. Yet, here a situation more akin
to the saturable analogue of the DNLS occurs~\cite{kip,champ},
whereby exchanges of stability between the on-site and inter-site
solutions arise. Likely, and analogously to corresponding DNLS
cubic-quintic or saturable settings, these exchanges are mediated
by pitchfork bifurcations (and reverse
pitchforks) generating asymmetric waveforms, a topic potentially
worthy of further investigation in the future. Additionally,
these states appear to possess quartet of eigenfrequencies
chiefly responsible for their instability. While this instability
appears to reach an asymptotic growth rate (over $\epsilon$ variations)
for values of $\Lambda$ below a critical one, it is an open problem
whether indeed this instability is expected to be present in the
continuum limit of the problem. In that connection, it is relevant
to point out that we have observed the manifestation of the instability
to potentially lead to traveling and mobility of the structure, while in other cases,
we have observed it to lead to a fragmentation of the solitary
wave into the staggered structures, a feature
which would not be ``accessible'' in the continuum limit.
It should also be pointed out that despite our computation for
different domain sizes as a function of $\Lambda$ for large
(approaching the continuum) values of $\epsilon$, the concavity
of the relevant eigenvalue dependence precludes a straightforward
determination of the associated critical value of $\Lambda$.
It can be safely inferred that the instability (at least for
sub-critical exponents $k<2/n$)
is absent in the nonrelativistic Schr{\"o}dinger limit.
Nevertheless, whether this occurs asymptotically as $\Lambda
\rightarrow 1$ or at a finite $\Lambda_c$ (the latter being observed
in the case of our finite -but large- coupling and domain size)
remains yet another important open question.

Naturally, the present investigation, as a primary one of its
kind, raises a considerable
volume of additional questions meriting
future examination both at the discrete and at the continuum limit.
In particular, a key issue is how
the discrete model asymptotes to the actual corresponding continuum.
It is especially important to understand how the spectral
properties may be modified in the limit. Another very interesting
avenue of research would be to develop and utilize
the solvability conditions that
were especially handy in the DNLS case to understand the unusual
existence and stability features in the corresponding Dirac case.
Understanding also better the role (especially in the dynamics)
of the unusual staggered structure would be especially relevant.
Other themes, such as a classification of the dynamical instability
scenaria for different states or the identification of the exponential
instability in the near-linear limit of small $k$ have also
emerged.
Beyond the realm of ``single pulses'' focused upon herein,
the cases of multi-pulses, pulse interactions and related themes are entirely
open, to the best of our knowledge, not only in the discrete case
but largely also in the continuum one. Finally, all these investigations
could naturally be generalized to higher dimensions, where also
vortical and related structures could potentially arise~\cite{pgk,pelinov}.
Some of these topics are currently under investigation and will
be reported in future publications.

\section*{Acknowledgements}

This work was supported in part by the U.S. Department of Energy (A.S.).
P.G.K. acknowledges support from the National Science
Foundation under grants CMMI-1000337, DMS-1312856, from FP7-People
under grant IRSES-606096 from
the Binational (US-Israel) Science Foundation through grant 2010239, and from
the US-AFOSR under grant FA9550-12-10332. We are indebted to Faustino Palmero for technical assistance with some parts of the manuscript.

\end{document}